\documentclass[pr, twocolumn, preprintnumbers, showpacs, footnoteadded, superscriptaddress,longbibliography]{revtex4-1}
\pdfoutput=1

\usepackage[T1]{fontenc}
\usepackage{amssymb}
\usepackage{graphicx}
\usepackage{amsmath}
\usepackage{slashed}
\usepackage{tensor}
\usepackage{hyperref}
\usepackage{epstopdf}
\usepackage{extarrows}
\usepackage{xcolor}

\newcommand{\td}{\text{d}}
\newcommand{\te}{\text{e}}

\newcommand{\be}{\begin{equation}}
\newcommand{\ee}{\end{equation}}
\newcommand{\bea}{\setlength\arraycolsep{2pt} \begin{eqnarray}}
\newcommand{\eea}{\end{eqnarray}}
\newcommand{\nn}{\nonumber}

\newcommand{\bpm}{\begin{pmatrix}}
\newcommand{\epm}{\end{pmatrix}}

\usepackage{changes}
\usepackage{braket}
\definechangesauthor[name={Per cusse}, color=orange]{per}

\begin{document}
\title{\boldmath  Using black holes  as rechargeable batteries and nuclear reactors}
\author{Zhan-Feng Mai}
\email{zhanfeng.mai@gmail.com}
\affiliation{Center for Joint Quantum Studies and Department of Physics, School of Science, Tianjin University, Yaguan Road 135, Jinnan District, 300350 Tianjin, P.~R.~China}
\author{Run-Qiu Yang}
\email{aqiu@tju.edu.cn}
\affiliation{Center for Joint Quantum Studies and Department of Physics, School of Science, Tianjin University, Yaguan Road 135, Jinnan District, 300350 Tianjin, P.~R.~China}

\begin{abstract}
This paper proposes physical processes to use a Schwarzschild black hole as a rechargeable battery and nuclear reactor. As a rechargeable battery, it can at most transform 25\% of input mass into available electric energy in a controllable and slow way. We study its internal resistance, efficiency of discharging, maximum output power, cycle life and totally available energy. As a nuclear reactor, it realizes an effective nuclear reaction ``$\alpha$ particles+black hole$\rightarrow$positrions+black hole'' and can transform 25\% mass of $\alpha$-particle into the kinetic energy of positrons. This process amplifies the available kinetic energy of natural decay hundreds of times. Since some tiny sized primordial black holes are suspected to have an appreciable density in dark matters, the result of this paper implies that such black-hole-originated dark matters can be used as reactors to supply energy.
\end{abstract}

\maketitle
\flushbottom

\noindent

\section{Introduction}
A basic property of a battery (the ``battery'' in this paper refers to all devices or systems that can generate electric energy) is that it can convert non-electric energy into electric energy in a controllable way to supply the appliances. There are several common mechanisms of battery, such as using the chemical energy, solar energy, wind or water energy, nuclear energy and so on. The black hole, being predicted by Einstein's general relativity, is one of the most fantastic object in our universe, describing an extreme region from which the gravity is so strong that no signal or classical matter can escape. Taking the fact that the black hole has extremely strong gravitational force, an interesting question arises: considering at least theoretically, could we use the gravitational force of black holes to generate electric energy, i.e. make use of black holes as batteries?

Though the black hole's strong gravity forbids that the classical matters escape from it into outside, fortunately, the energy can be extracted from the black hole through quantum or classical processes. The famous Hawking radiation is one quantum way that the black hole losses energy~\cite{Hawking2,Hawking3}. For a larger black hole such effect is too weak to be used. For a small enough black hole, the  Hawking radiation can be strong. However, in this case, the black hole will evaporate in a very short time and the energy releasing is too drastic and rapid to be used as a controllable supply. An other of the famous processes for energy extraction is the Penrose process, stating that a classical particle can extract rotation energy from Kerr black hole~\cite{penrose1971extraction,chandrasekhar1998mathematical}. Furthermore, it also turned out that there exist an effective Penrose Process arising between the Reissner-Nordstr\"{o}m (RN) black hole and a charged particle, showing that the energy and charge of RN black hole can be extracted by charge particles~\cite{denardo1973energetics}. In addition, another classical process being able to extract energy from black holes is superradiance~\cite{Brito:2015oca}, which can also extract energy from rotational or charged black holes.

By Penrose process or superradiance, the charged black hole can be used as a kind of battery. To extract the electric energy of same order as the black hole mass, the black hole must be almost extreme, i.e. contains a large amounts of charge. However, the naturally formed black holes, which come from the primordial black hole, the large mass star or the other astrophysical processes, though may carry large amounts angular momentum, will be neutral.  The observations via gravitational wave~\cite{PhysRevLett.116.061102,PhysRevLett.118.221101} also verified this point. Particularly, the Schwarzschild black hole is the final stable state of black hole in classical physics and so there is seemingly no hope to use it as a battery to supply appliances by classical processes.

In this paper we propose a method to recharge Schwarzschild black holes by very tiny amounts of charge. When the radius of initial black hole is large enough, we show that strong gravitational force can transform at most 25\% input mass into electric energy.  We show that the Joule heat produced in discharging process just satisfies the requirement of thermodynamical laws of black holes. We also study the discharging efficiency, maximum output power, cycle life and  totally available energy. By making use Schwinger effect, we show that the tiny black hole can be used as a reactor to transform the $\alpha$-particles into positrons. This black hole reactor may amplify the kinetic energy of $\alpha$-decay hundreds of times. Interestingly, we find that the black hole mass of such reactor just locates inside the window that black hole could be a candidate of dark matters.

\section{Add fuel for battery}
Let us first introduce how to add ``fuel'' (recharging process) for a Schwarzschild black hole. In following discussion, we mainly use the natural units system $G=c=\hbar=k_B=4\pi\varepsilon_0=1$.

\begin{figure}
  \begin{center}
\includegraphics[width=.15\textwidth]{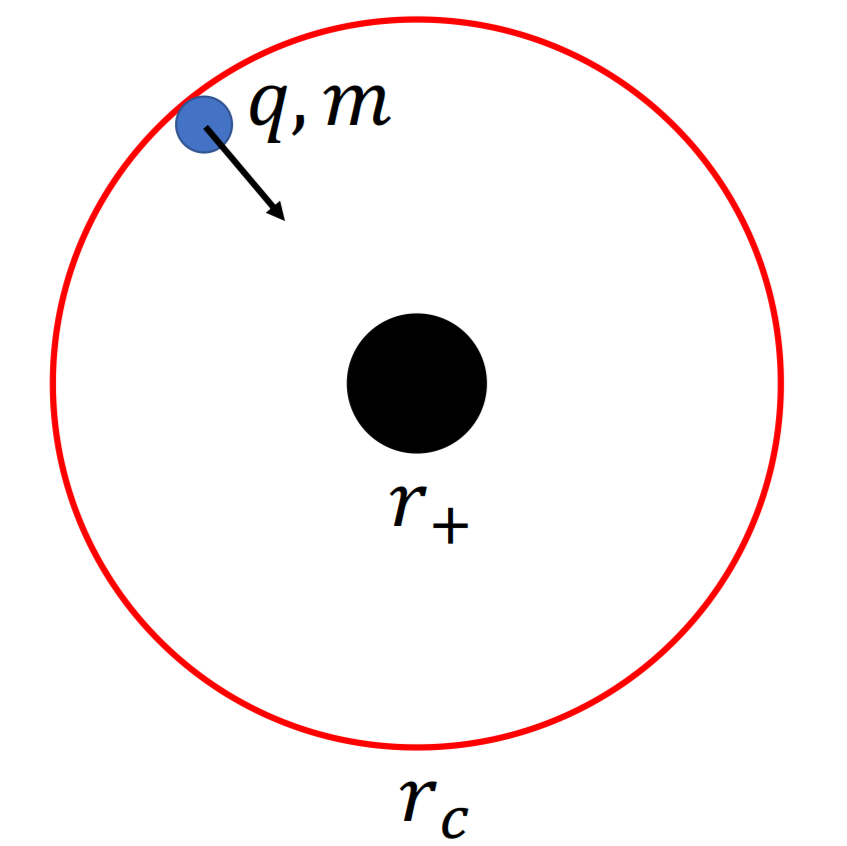}
  \caption{The charged sphere falls into black hole spontaneously since the gravitational force is stronger than Coulomb repulsion.}\label{recharge1}
  \end{center}
\end{figure}
Consider a static spherically symmetric black hole has mass $M$ and charge $Q$ with $|Q|/M\ll1$, which is confined inside a cavity. We assume that the radius $r_c$ of cavity is much larger than the horizon radius $r_+$ of the black hole. The black hole cannot be too small since the Hawking radiation will lead that a tiny black hole evaporates completely in a very short time. If we require that effect of Hawking radiation is negligible in a time scale of age of our universe, the minimal mass of initial black hole should be much larger than $10^{11}$kg. We prepare a large amounts of charged particles, every one of which contains the same mass $m$ and charge $q$ with
\begin{equation}\label{relqm1}
 |q|\gg m,~~\text{or in SI unit }|q|/m\gg10^{-10}\mathrm{C/kg}
\end{equation}
If we inject a such sphere into the cavity (see Fig.~\ref{recharge1}) under following restriction
\begin{equation}\label{requiren1}
 \frac{Q}{M}q\leq m,
\end{equation}
then gravitational force will always larger than electric repulsion and these spheres will spontaneously fall into black hole. This result can be obtained from Newton's laws combining with Coulomb's formula. Interestingly, the general relativity will also give us same result, see Sec.~I of Supplementary Materials for details. By this process, the neutral black hole is recharged. Note that this recharging process does not consume additional energy since the falling of charged particles is spontaneous.

On the other hand, when the electric field $E$ near horizon is strong enough, the Schwinger effect may play an important role~\cite{PhysRev.82.664}. In this case the electron-positron are separated from vacuum and the charged black hole will absorb one of them, which prevents the black hole from being recharged further. We note the Schwinger effect will happen at a scale of electron Broglie wavelength $\lambda_e\approx 10^{-12}$m. In the case that the black hole horizon radius $r_+\gg\lambda_e$ (this will be the case considered in this paper), the equivalent principle then implies that gravitational effect can be ignored when we consider the physics below the length scale $\lambda_e$. We then obtain following inequality
\begin{equation}\label{Schwinger1}
  \frac{|Q|}{r_+^2}\approx\frac{|Q|}{4M^2}\lesssim E_s\,.
\end{equation}
Here $E_s\approx10^{18}$V/m is the critical field of Schwinger effect. If inequality~\eqref{Schwinger1} is broken, the black hole can absorb opposite charge from vacuum and the battery cannot be charged anymore.

When any one of conditions~\eqref{requiren1} and \eqref{Schwinger1} is broken, the recharged process will stop. Assume that an initially neutral black hole has mass $M_0$ and can at most swallow $n_r$ such charged particles. The finally charge and mass of fully recharged black hole then read $Q_r=n_rq$ and $M=M_0+n_rm$. If the inequality~\eqref{requiren1} is first broken, which we call ``Coulomb domain'', the charged particle cannot fall into black hole spontaneously since Coulomb repulsion will be stronger than gravitational attraction. In this case we have $n_r=M_0m/q^2$. If the inequality~\eqref{Schwinger1} is first broken, which we call ``Schwinger domain'', the Schwinger effect prevents the black hole from being recharged and we have $n_r=4E_sM_0^2/|q|$. They can be combined into
\begin{equation}\label{numberq}
  n_r=\min\left\{\frac{M_0m}{q^2},~~\frac{4E_sM_0^2}{|q|}\right\}\,.
\end{equation}
or in SI units
\begin{equation}\label{numberq2}
  n_r=\min\left\{\frac{M_0m/\mathrm{kg}^2}{(q/\mathrm{C})^2}\times10^{-20},~\frac{(M_0/\mathrm{kg})^2}{|q/\mathrm{C}|}\times10^{-46}\right\}\,.
\end{equation}
The ``Coulomb domain'' will happen when $M_0$ is large enough; while the  ``Schwinger domain'' will happen for small-sized black holes.

From Eqs.~\eqref{numberq} and \eqref{relqm1} we see that the $Q_r$ and increased mass $\Delta M$ of fully recharged black hole satisfies
\begin{equation}\label{qrur1}
  |Q_r|=n_r|q|\leq\frac{M_0m}{q^2}|q|=\frac{m}{|q|}M_0\ll M_0
\end{equation}
and
\begin{equation}\label{qrur2}
  \Delta M=n_rm\leq\frac{m^2}{q^2}M_0\ll M_0\,.
\end{equation}
%
The horizon radius of fully recharged black hole becomes
\begin{equation}\label{rplus1}
  r_+=M+\sqrt{M^2-Q_r^2}\approx2M_0+2n_rm-\frac{n_r^2q^2}{2M_0}\,.
\end{equation}
The Hawking temperature reads $T\approx1/(8\pi M_0)$. The increased energy $\Delta M$ contains two parts, the first one causes the increment of entropy~\cite{PhysRevD.7.2333,PhysRevD.9.3292,Hawking2,Bardeen1973,hawking1}, which stands the thermal dissipation $T\Delta S\approx\Delta r_+/2$. This is the unavailable energy.  The other one
\begin{equation}\label{deltaM1}
  \mathcal{E}=\Delta M - T \Delta S \approx\frac{n_r^2q^2}{4M_0}=\min\left\{\frac{n_rm}4, M_0^3E_s^2\right\}
\end{equation}
is available energy, stored as the static electric potential energy. Furthermore, the efficiency for the recharging process is $\eta_{r} =\mathcal{E}/(n_rm)\leq25\%$, i.e. black hole can at most transform 25\% input mass into available energy of static electric field.

 After adding fuel for battery, let us discuss which black hole is good potential candidate for our black hole batter. It is widely known that the stellar black hole with solar mass is surrounded by interstellar matter, such as plasma, accretion disk et al.  The stellar black hole will thus discharge by surrounded matter very quickly. 
	
However, for tiny black hole, especially in atomic scale, most of matter around the black hole will be captured and absorbed very quickly. For example, one can see that the damping time scale of bosonic as well as Fermionic field near the tiny black hole from the associated quasinormal modes $\Gamma \sim 10^{-16}{\rm s}$~\cite{Berti:2009kk,Konoplya:2011qq}. Any fundamental particle around this tiny black hole will either escape into infinite or be absorbed very quickly. Therefore, our potential black hole battery should be tiny black hole in atomic scales. Such a small black hole most likely be primordial black hole, which is a candidate of dark matter in our Universe and may exist widely. We thus propose rechargeable battery and nuclear reactor using classical Schwarzschild black hole with $10^{15} {\rm kg} \sim 10^{18} {\rm kg}$.  In such small scale, the accretion disk around our tiny black hole cannot form ~\cite{Bambi:2008hp}.

\section{discharge process}\label{theory}
\subsection{Effective circuit}
We now consider the discharge process. As a toy model, we open a complex massless scalar field $\Psi$ in the box to minimally couples with gravity. The complex scalar field carries charge $\sigma$ and is confined in the box. In frequency domain~\cite{broughton2018}, we consider the s-wave mode approximation and adopt the Dirichlet boundary condition for the scalar field
\begin{equation}\label{source}
\Psi = \te^{-i \omega t} {\cal R}(r)/r \, , \quad {\cal R}|_{r=r_c}= X_0,~~\omega\geq0\, .
\end{equation}
We further leave $\omega$ and $X_0$ as two parameters that we can control. Our discharging process indeed makes use of superradiance, which treats the complex scalar as perturbation, by neglecting the backreaction of scalar field on spacetime geometry. A detailed study on the dynamics of black hole in cavity will be found in Sec.~II of Supplementary Materials and relative Refs.~\cite{PhysRevD.47.1407,BROWN2002175,Szabados2004,Maldacena:1997re,Gubser:1998bc,Witten:1998qj,Witten:1998zw,Ishii:2022lwc}. This in fact is ``quasi-static approximation'' and can be applied when ${\cal R}(r)$ is small enough. The theory of superradiance gives us following energy flux density ${\cal P}_{\rm e}$ and charge current density ${\cal P}_{\rm Q}$ on the boundary of the box~\cite{Brito:2015oca}
\begin{eqnarray}\label{power1a}
 &&{\cal P}_{\rm e} =\frac{2\sigma \omega}{r_c^2} \left( \frac{\omega}{\sigma}-\mu_h \right)|X_0|^2|{\cal T}_h|^2 \, , \cr
 && \\
 &&{\cal P}_{\rm Q} = \frac{2\sigma^2}{r_c^2}\left(\frac{\omega}{\sigma} - \mu_h \right)|X_0|^2|{\cal T}_h|^2 \, . \nonumber
\end{eqnarray}
Here we denote $\mu_h=Q/r_+$ as the electric potential on the horizon and ${\cal T}_h$ as the transition amplitude for the complex scalar near the horizon. In this approximation the discharging process in indeed makes use of superradance. It is thus nature to define the electric current $I$ and power $P_d$ in an effective circuit as
\begin{equation}\label{current}
I =-4\pi r_c^2 {\cal P}_{\rm Q},~~P_d=-4\pi r_c^2{\cal P}_{\rm e}\, .
\end{equation}
Furthermore, we can define the terminal voltage $U_d=\omega/\sigma$ and the electromotive force $U_{\rm bh}=\mu_h$. The Eq.~\eqref{power1a} can therefore be written as the Ohm's Law
\begin{equation}\label{Ohm}
P_d =U_dI \, , \quad (U_{\rm bh}-U_d)=I R \, ,
\end{equation}
where the effective internal resistance $R$ of the battery reads
\begin{equation}\label{resis}
R \equiv\frac1{ 8\pi \sigma^2 |{\cal T}_h|^2|X_0|^2} \, .
\end{equation}
Due to the existence of internal resistance,  the Joule heat will be created, which stands for the unavailable work from the battery and will cause the increment of entropy. Thus, the discharging is quasi-static but irreversible process.

Since the system is quasi-static,  the first law of the thermodynamics for the RN black hole shows
\begin{equation}
T \frac{\td S}{\td t} = \frac{\td M}{\td t} - \mu_h \frac{\td Q}{\td t}=-P_d+\mu_hI \, ,
\end{equation}
where $T$ and $S$ denote the Hawking temperature and entropy of the RN black hole respectively. We thus have
\begin{equation}\label{entropy}
T \frac{\td S}{\td t}=8\pi \sigma^2\left(\frac{\omega}{\sigma}-\mu_h\right)^2 |X_0|^2 |{\cal T}_h|^2\, .
\end{equation}
We see that, in discharging process, the entropy of our black hole battery is always increasing, following the requirement of  Hawking's area theorem of black hole horizon. Furthermore, more importantly, from Eqs.~\eqref{current}, \eqref{resis} and \eqref{entropy}, one will easily find $T\td S = I^2 R \td t$. This shows that the increment of entropy exactly corresponds to the heat dissipation produced from the effective internal resistance of our battery. 


As a device of energy supply, the efficiency and power are main parameters to characterize its performance. Due to the nonzero the effective internal resistance, the  maximal efficiency of discharging will happen when the current $I\rightarrow0$. In this case, the thermal dispersion caused by internal resistance is negligible and our battery can at most transform 25\% input mass into available electric energy. Noting the fact that the an atom bomb only releases 0.1\% energy of its mass. The maximal efficiency of the battery is 250 times higher than that of the atomic bomb.

An other interesting question, and maybe more more practical concern, is the maximum power and its corresponding efficiency~\cite{Curzon1975,PhysRevLett.95.190602,PhysRevLett.102.130602}. In order to simplify the discussion, we will apply an additional approximation named ``high frequency approximation'': $\omega M\gg1$. Then transmission amplitude can be approximated by $|{\cal T}_h|^2 \approx 1$. This will mean that the effective resistance becomes constant, namely $R_0 \equiv R|_{\omega M \gg 1} \approx 1/(8\pi \sigma^2|X_0|^2)$. It is from Eq.~\eqref{power1a} easy to observe that there exist an instantaneous maximal output power when the discharging voltage is half of the electromotive force at every moment $t$, i.e. $\omega/\sigma=\mu_h/2$.  The battery can transform 12.5\% input mass into available energy if it works at maximal power.

To end this subsection, we shall discuss the problem of superradiant instability of the charged scalar field coupled with the charged black hole in a cavity~\cite{Herdeiro:2013pia,Sanchis-Gual:2015lje,Brito:2015oca}.  In our black hole battery, we set the complex scalar with time-dependent exterior source $\braket{\cal O}$ on the boundary of the cavity (See also Appendix.~\ref{dycavity}). It is similar to the mechanical system with forced vibration. The state of absorbing or releasing energy for such a system depends on the frequency of the external source. Therefore, the charge and discharge process of our black hole battery depends on the source, $\braket{\cal O}$ on the cavity boundary.
	
Moreover, the superradiant instability of a charged black hole in the cavity will only happen when the cavity is a mirror. In other words, the system is an "autonomous system", where the complex scalar has no time-dependent exterior source. Therefore, the superradiant instability will not happen in our black hole battery model during the discharging process.

\subsection{Cycle life and totally available energy at maximum power supply}
For a neutral black hole confined in a cavity, we can use above mentioned method to recharge it fully and then discharge it completely. Such process can be repeated many times and so this battery is rechargeable battery.
After a single recharging---discharging cycle, even though the black hole battery goes from Schwarzschild black hole to Schwarzschild black hole again, the entropy increase. The Hawking-Bekenstein entropy formula shows that the Schwarzschild radius of our black hole battery will always increase after every such cycle. Consequently, when the radius is near the boundary $r_c$, our black hole battery become unavailable.

Here we consider the case that the exterior black hole is large enough so the recharging process will stop at ``Coulomb domain''. We first obtain the increment of the Schwarzschild radius for the recharging process $\Delta r_+=(n_r^2q^2/4M_0^2)r_+=(m^2/4q^2)r_+$ from Eqs.~\eqref{rplus1} and \eqref{numberq}.  To estimate maximal cycle life, we use the minimal power to discharge, i.e. $I\rightarrow0$, so that the entropy increment in discharging process is negligible.  Therefore,
\begin{equation}\label{radiusn1}
  r_{+,n}=\left(1+\frac{m^2}{4q^2} \right)r_{+,n-1}\,.
\end{equation}
Here $r_{+,n}$ stands for the black hole radius after $n$ times recharge---discharge cycle process. It implies that
\begin{equation}
r_{+,n} =2M_0\left(1+\frac{m^2}{4q^2}\right)^n \, ,
\end{equation}
where we have imposed $r_{+,0} = 2M_0 $ as the initial Schwarzschild radius of our black hole battery. Furthermore, when $r_{+,N} = r_c/\alpha$ with truncation $\alpha\gtrsim1$, our horizon is close to the boundary and the black hole battery's life comes to the end. The maximal number of the recharge---discharge cycle process then reads
\begin{equation}\label{life0}
N_{\text{max}}\approx\frac{\ln (r_c/\alpha) - \ln 2 M_0}{\ln (1+m^2/4q^2)}\approx\frac{4q^2}{m^2}\ln\left(\frac{r_c}{2 M_0\alpha}\right)\, .
\end{equation}
We then obtain an estimation on the maximal total available energy of its whole life $E_{\rm total, max }\sim N_{\text{max}}\mathcal{E}$. Since $\mathcal{E}= n_rm/4$ and $n_r= M_0m/q^2$ at ``Coulomb domain'', we then have
\begin{equation}
E_{\rm total, max }\sim M_0\ln\left(\frac{r_c}{2 M_0\alpha}\right) {\color{red} >} M_0\, .
\end{equation}
The black hole battery during its life can even supply the energy larger than its initial mass.

\begin{figure}
  \begin{center}
\includegraphics[width=.35\textwidth]{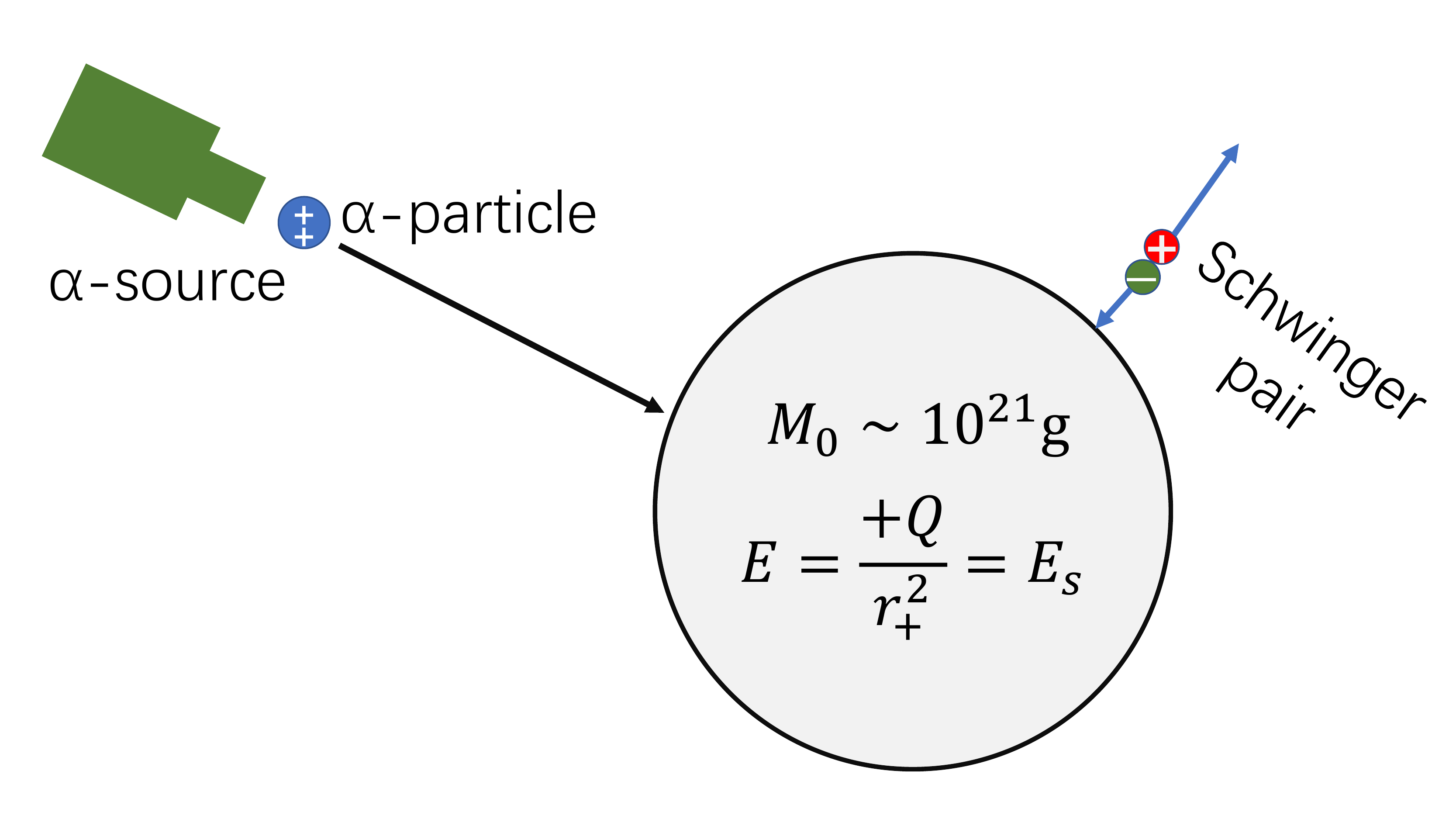}
  \caption{At Schwinger domain the black hole spontaneously absorbs the $\alpha$-particles and radiates positrons.} \label{reactor1}
  \end{center}
\end{figure}
%
As we illustrated above, our black hole battery is based on the energy and charge extraction by superradiance. There is another similar process for extracting energy from black hole, called the Blanford-Znajek process~\cite{Blandford:1977ds}. Even though both superradiance and the mentioned Blanford-Znajek process, described by circuit-theory analogy, can achieve the energy extraction process, the physics are different. 

Firstly, our black hole battery model focuses on extracting the charge with energy from a tiny charged Schwarzschild black hole (RN black hole) by perturbed complex scalar with source. However, the Blanford-Znajek process extracts the magnetic energy from the supermassive rotating black hole with an accretion disk. Secondly, the Blanford-Znajek process is widely known and play an important role in astrophysical black hole, especially stellar black hole. Nevertheless, as we mentioned above,the potential candidate for our black hole battery is a tiny spherical black hole in an atomic scale, with which the accretion disk is difficult to form. Therefore, the mechanism of energy extraction, as well as the environment of our black hole battery model and Blanford-Znajek process are different.

\section{Black hole reactor}
From nuclear physics, we know that the reaction $\alpha\rightarrow\mathrm{e}^+$ is forbidden in particle physics due to the conservation of the baryon number. However, we in the following will propose a process to spontaneously realize an effective reaction by taking a black hole into account.

The basic idea is shown in Fig.~\ref{reactor1}. We prepare a natural $\alpha$-source at the safe distance far away from the black hole. The $\alpha$-particles has $q=2e$ and $m=m_{\alpha}\approx4$GeV with typical kinetic energy $\mathcal{E}_k\approx 5$MeV.  The kinetic energy is much smaller than its rest mass and so is negligible.  Then we use above mentioned process to recharge the neutral black hole. Initially the strong gravity swallows the $\alpha$-particle into black hole. We assume that the initial mass of black hole satisfies
\begin{equation}\label{Meoes1}
  10^{15}\mathrm{kg}\ll M_0\leq m_\alpha/(4E_sq)\approx10^{18}\mathrm{kg}\,.
\end{equation}
In this mass region, the horizon radius of black hole will be much larger than the electron Broglie wavelength $10^{-12}$m and the $\alpha$-particle radius $10^{-15}$m. The electrons, positrons and $\alpha$-particles can all be treated as point-particles. We see that the recharging process will be terminated at Schwinger domain. To recharge the initial black hole into critical field only needs very tiny amount of $\alpha$-particles $n_r\lesssim10^{10}$. When Schwinger critical electric field arrives, the black hole will still spontaneously swallow the electrons but ``eject'' the positrons from vacuum. The electric field of black hole will accelerate the positrons and energy of positrons that escape into infinity can be estimated by
\begin{equation}\label{energye}
  \mathcal{E}_p\sim E_ser_+\approx2E_sM_0e\leq\frac{m_{\alpha}}4\,.
\end{equation}
We see that if the initial mass of black hole near $10^{18}$kg, the energy of positron will approach to 1GeV. Compared with the kinetic energy of input $\alpha$-particles, the kinetic energy of positron is amplified hundreds of times.

Such tiny black holes are primordial black holes, which are formed at early universe and can exist till today. Interestingly,
as we mentioned above, they are just suggested to have an appreciable density as candidate of dark matters~\cite{Carr:2020xqk}. If this is true, such size black hole should widely exist in our universe and we find a new application for dark matters: it may play a role of antimatter factory~\cite{Bambi:2008hp} and can be used as an efficient reactor to amplify MeV $\alpha$-radiation into GeV positron radiation.

Even though we share similar applications on the Schwinger effect to produce anti-particles with Ref.~~\cite{Bambi:2008hp}, the purposes are different. The Ref.~~\cite{Bambi:2008hp} suppose that the black hole with low mass (around $10^{21} {\rm g}$ ) can be surrounded by plasma. The protons in the plasma then can be absorbed by the black hole. Then, once the electronic field around the black hole is strong enough, the Schwinger effect happens, emitting positrons.
	
As we illustrated above, matters is difficult to surround the tiny black hole in atomic scale. we then provide another scheme to realize this process, building an effective nuclear reactor using tiny black hole. For example, after obtaining the $\alpha$ particle using \emph{prepared $\alpha$ source}, we inject the $\alpha$ particles into charged black hole to trigger the Schwinger effect to produce positrons.  Using this effect, together with the black hole, we realize an effect  ``$\alpha$ to $e^{+}$'' reaction, which is forbidden in nuclear physics. Furthermore, our black hole reactor model realizes the transformation from MeV $\alpha$ particle into GeV radiation, which greatly improves the efficiency of nuclear reaction.

\section{Conclusion}
In this paper, we argue theoretically that we can use a Schwarzschild black hole as a rechargeable battery. We propose a way to add ``fuel'' for this battery (recharging process) and show that it at most can transform 25\% input mass into electric energy. We show that the recharging and discharging processes are always irreversible processes, as required by the second law of black hole physics. We then further identify the effective internal resistance in discharging process and show that the Joule heat produced by the effective resistance is just as same as the amounts required by the laws of black hole thermodynamics. We study the discharging efficiency, maximum output power, cycle life and totally available energy. Using tiny black holes, we propose a model of black hole reactor, which realizes an effective nuclear reaction ``$\alpha+$black hole$\rightarrow\mathrm{e}^++$black hole''. This process can at most amplifies the kinetic energy of $\alpha$-decay hundreds of times. Interestingly, the optimal mass of such a black hole reactor is just located inside the window of primordial-black-hole-originated dark matters.

\begin{acknowledgments}
We thank Song-Song Pan to help us correct typos of this manuscript. This work is supported by the National Natural Science Foundation of China under Grant No. 12005155.
\end{acknowledgments}

\appendix
\section{The Condition of mass-charge ratio for dropped charged Particles}\label{app0}
We consider a classical charged particle with mass $m$ and charge $q$ moving along the geodesic in the RN black hole background. The action for this particle in spherical coordinate $x^\mu = (t,r,\theta,\varphi)$ is
\begin{equation}
S = \int \td \tau L \, , \quad L =  \frac{1}{2\lambda}g_{\mu\nu}\dot{x}^\mu \dot{x}^\nu - \frac{\lambda}{2}m^2 -q \dot{x}^\mu A_\mu
\end{equation}
where $\tau$ denotes the affine coefficient and $\dot{x}^\mu \equiv \frac{\td x^\mu}{\td \tau}$. $\lambda$ is an auxiliary field, performing variation respect to which gives the normalization of 4-velocity that
\begin{equation}
g_{\mu\nu}\dot{x}^\mu \dot{x}^\nu = -1 \, .
\end{equation}
In the background of RN black hole with mass $M$ and charge $Q$, namely
\begin{equation}\label{RNmetric}
\td s^2 = -f \td t^2 + \frac{\td r^2}{f} + r^2 \td \Omega^2_2 \, , \quad  A_t = \frac{Q}{r} \,
\end{equation}
and $f=1-2M/r + Q^2/r^2$, if we consider a timelike killing vector $\chi^\mu = (\partial/\partial t)^\mu$, the charged particle's action will be invariant under the infinitesimal transformation $x^\mu \to x^\mu + \epsilon \chi^\mu\, , \quad \epsilon \ll 1$, namely
\begin{equation}
{\cal L}_\chi S = -\int \td \tau q \dot{x}^\mu (\chi^\nu \partial_\nu A_\mu + A_\nu \partial_\mu \chi^\nu) = 0 \, ,
\end{equation}
where we have used the Killing equation $\partial_{(\mu} \chi_{\nu)} = 0$, as well as the fact that ${\cal L}_\chi g_{\mu\nu} = 0 $ since the static RN black hole is not dependent on $t$. It will give the following conserved energy of the massive charged particle,
\begin{equation}\label{dtdtau}
E \equiv \left(1-\frac{2M}{r}+\frac{Q^2}{r^2}\right)\frac{\td t}{\td \tau} + \frac{q Q}{m r}\, .
\end{equation}
Since in the recharging process we have $qQ\geq0$, we see $E>0$. Furthermore, for simplicity, we consider the initially static particle without angular momenta, lying on the plane $\theta = \pi/2$. Recall the normalization of 4-velocity. we obtain the radial worldline of this particle as
\begin{equation}
\left(\frac{\td r}{\td \tau}\right)^2 = E^2 - V_{\rm eff}\, ,
\end{equation}
where the effective potential $V_{\rm eff}$ is
\begin{equation}
V_{\rm eff} = 1 - \frac{2M}{r}+\frac{2 E q Q}{m r} - \left(\frac{q^2}{m^2}-1\right)\frac{Q^2}{r^2}  \, .
\end{equation}
In asymptotic flat region, if the particle is relaxed freely at infinity then we have $E=1$. If it falls into the RN black hole, namely the total interaction, including gravitational and electrical one, applied to charged particles are attractive, it is required that
\begin{equation}
V_{\rm eff} |_{r \to -\infty}  \approx 1 + \frac{\alpha}{r} + {\cal O}\left(\frac{1}{r^2}\right) \,  ,\quad \alpha < 0 \, .
\end{equation}
This indicates that
\begin{equation}\label{mmqq1}
\frac{m}{q} \geq \frac{Q}{M} \, .
\end{equation}
Under our requirement $|q|/m\gg1$, we see that condition~\eqref{mmqq1} then insures
\begin{equation}\label{EminusV}
  \left(\frac{\td r}{\td \tau}\right)^2 =\frac{2M-2Qq/m}r+\left(\frac{q^2}{m^2}-1\right)\frac{Q^2}{r^2}>0
\end{equation}
This means that $\td r/\td\tau<0$ outside horizon. Eq.~\eqref{dtdtau} shows
\begin{equation}\label{dtdtau1}
  \frac{\td t}{\td\tau}=\left(1-\frac{Qq}{mr}\right)/f\geq\left(1-\frac{M}{r}\right)/f>\frac1{2f}\,,
\end{equation}
so we find $\td r/\td t\leq0$ outside black hole. Thus, the particle will fall from infinity into the horizon without backtracking.

Intriguingly, in context of Newtonian mechanic, this condition also can be given that the gravitational attraction applied to the charged particles is larger than the electrical repulsion, leading to these particles falling into black hole.

\section{Dynamics of black hole in cavity}\label{dycavity}

\subsection{On-shell Action and Ward Identities}
We begin with the Einstein-Maxwell theory, which minimally coupled a free massless complex scalar field.  The complex scalar field carries charge $\sigma$.  In this paper, we will always assume that the quantum effects of both black holes and matters are negligible.
The total action of our theory is written as
\begin{equation}\label{action}
S = S_{\rm bulk} + S_{\rm surf}
\end{equation}
where
\begin{eqnarray}\label{action2}
&& S_{\rm bulk} := \frac{1}{16\pi}\int_M \td^4 x \sqrt{-g} \left(R - F^2 -|D_\mu \Psi|^2 \right)\, ,  \\
&& S_{\rm surf} :=-\frac{1}{8\pi} \int_{\partial M} \td^3 x \sqrt{-h} ( K - K_0 ) \,.
\end{eqnarray}
In the bulk action, $F_{\mu\nu} \equiv \partial_\mu A_\nu - \partial_\nu A_\mu$ stands for the field strength of the Maxwell field, as well as $D_\mu := \nabla_\mu - i \sigma A_\mu$. The surface action includes the Gibbons-Hawking term and the counter term respectively, where $h_{ij} $ and $K$ denote the induce metric the associated intrinsic curvature scalar. We here choose $K_0 = K|_{g_{\mu\nu} = \eta_{\mu\nu}}$ to eliminate the infinity of flat spacetime.
Performing variation with respect to $g^{\mu\nu}$, $A_\mu$ and $\Psi^{\dag}$, the equations of motion read
\begin{eqnarray}\label{eom}
&&G_{\mu\nu}=8\pi T_{\mu\nu},~\nabla^\mu F_{\mu\nu}=-4\pi J_\nu^{\Psi},~D^\mu D_\mu \Psi =0\,,
\end{eqnarray}
where $G_{\mu\nu}=R_{\mu\nu}-\frac{1}{2}g_{\mu\nu}R$ denotes the Einstein tensor. The bulk energy momentum tensor $T_{\mu\nu}$ and the bulk current $J^{\Psi}_\mu$ read
\begin{eqnarray}
&& T_{\mu\nu} :=\frac1{4\pi}\left(F_{\mu\sigma}{F_\nu{}^{\sigma}}-\frac14g_{\mu\nu}F_{\rho\sigma}F^{\rho\sigma}\right) \cr
&& +\frac1{16\pi}
\left(D_\mu\Psi (D_\nu\Psi)^\dagger+D_\nu\Psi (D_\mu\Psi)^\dagger-g_{\mu\nu}\left(|D_\rho \Psi|^2\right)\right)\, , \cr
&& \cr
&& J^{\Psi}_\mu  :=-\frac{\sigma}{4\pi} \Im(\Psi^\dagger D_\mu\Psi) \, .
\end{eqnarray}

One widely used quasi-local definition of energy is the Brown-York energy momentum tensor~\cite{BROWN2002175,Szabados2004,Skenderis:2000in} from the on-shell action
\begin{equation}
{\cal T}_{ij}(x)\equiv \frac{-2}{\sqrt{-h}}\frac{\delta S_{\rm on-shell}}{\delta h^{ij}(x)}= [K_{ij}]-h_{ij}[K] \, ,
\end{equation}
Here $h_{ij}$ and $K_{ij}$ are induced metric and extrinsic curvature of cut-off boundary, $[K_{ij}]=K_{ij}-K_{0,ij}$  and $[K]=K-K_0$. The $K_{0,ij}$ is the extrinsic curvature when the boundary embedded in flat spacetime. This can be regarded as the energy momentum tensor of the black hole confined in the box. Since the black hole is allowed to exchange energy through the boundary, the Brown-York energy momentum tensor ${\cal T}_{ij}$ is not conserved. The dynamics of the Brown-York energy momentum tensor are ruled by the following ``Ward Identity'', arising from the diffeomorphic invariance of cut-off boundary.

Specifically, let $\xi^i$ to be arbitrary vector field which is tangent to cut-boundary, this vector field then can generate an infinitesimal diffeomorphic transformation $x^i\rightarrow x^i+\varepsilon\xi^i$ for boundary. The diffeomorphic invariance of the on-shell action requires that.
\begin{eqnarray}\label{Liede}
{\cal L}_{\xi} S_{\rm on-shell} = \int_{r_c} && \td^3 x \sqrt{-h}  \left({\cal T}_{ij}{\cal L}_\xi h^{ij} + J^i (x) {\cal L}_\xi A_i \right.\cr
&& ~\\
 &&\left. +  {\cal O}(x) {\cal L}_\xi \Psi^\dagger + {\cal O^\dagger}(x) {\cal L}_\xi \Psi \right) = 0 \nn \, ,
\end{eqnarray}
where ${\cal L}_\xi$ denotes the Lie derivative along with $\xi^i$ and
\begin{eqnarray}\label{respon}
&&  O(x) \equiv  \frac{1}{\sqrt{-h}} \frac{\delta S_{\rm on-shell}}{\delta \Psi^\dagger(x)}  \, , \\
&&J^i(x)  \equiv  \frac{-1}{\sqrt{-h}}\frac{\delta S_{\rm on-shell}}{\delta A_i (x)} \, .
\end{eqnarray}
Here we call $O(x)$ and $J^i(x)$  to be the expectation value of the response for the source of the complex scalar and the 3-electromagnetic current respectively, following the usual definitions of holographic duality in asymptotically anti-de Sitter spacetime~[18-21]. Note a similar deduction has also been used to study superradiance of asymptotically anti-de Sitter black holes~[22].

After integrating by parts, the ``Ward Identity'' for the Brown-York energy momentum tensor can be derived from the Eq.~\eqref{Liede}
\begin{equation}\label{ward1}
\bar{\nabla}^j {\cal T}_{ij} = - 2 \Re(O\bar{\nabla}_i \Psi^\dag) + \bar{\nabla}_{j}J^jA_i + J^jF_{ij} \, .
\end{equation}
where $\bar{\nabla}_i$ denotes the covariant derivative operator for the induced metric.  Additionally, our theory has not only the diffeomorphic invariance but also the gauge symmetry. The gauge symmetry requires that the on-shell action $S_{\rm on-shell}$ is invariant under the following gauge transformation,
\begin{equation}\label{ward2}
A_i \to A_i + \partial_i \lambda \, , \quad \Psi \to \te^{i q \lambda}\Psi\, .
\end{equation}
Specially, for infinitesimal $\lambda$, the gauge invariance implies the second ``Ward Identity'' for the current $J^i$
\begin{equation}\label{ward21}
\bar{\nabla}_i J^i = 2 \sigma \Im \left(O  \Psi^\dagger\right) \, .
\end{equation}

We now consider the physical interpretations on these two ``Ward Identities''. When the complex scalar field vanishes, Eq.~\eqref{ward21} becomes $\bar{\nabla}_i J^i=0$. This just reflects the charge conservation interior the box due to the fact that cut-off boundary is insulated with black hole. We then denote
\begin{equation}\label{ward22}
{\cal P}_{\rm Q} \equiv 2 \sigma \Im \left( \Psi^\dag \braket{\cal O} \right) \, .
\end{equation}
According to Eq.~\eqref{ward21}, we interpret ${\cal P}_{\rm Q}$ as the `` charge current density'' via the boundary of the box.

To understand the physical meaning of quantities appearing in the right-side of Eq.~\eqref{ward1}, let us denote that $\chi^i = \left(\frac{\partial }{\partial t} \right)^i$ is a timelike vector which is tangent to boundary. Define the energy current $J_{\rm e}^i \equiv {\cal T}^i{}_j  \chi^j$ and electric field $E_i=\chi^jF_{ji}$ at the cut-off boundary, the first ``Ward Identity '', Eq.~\eqref{ward1} gives
\begin{equation}\label{ward11}
\bar{\nabla}_j  J^j_{\rm e} = {\cal T}^{ij}\bar{\nabla}_i\chi_j- 2 \Re(O \partial_t \Psi^\dag) + \mathcal{P}_QA_t + J^iE_i \, .
\end{equation}
The terms appearing in right-side of Eq.~\eqref{ward11} can be explained as follows: The $J^iE_i$ stands for the electric power caused by the interaction between electric field $E_i$ and current $J^i$ on the boundary; the  $\mathcal{P}_QA_t$ stands for the energy increasing caused by the charge increasing since $A_t$ stands for the chemical potential of charge and $\mathcal{P}_Q$ stands for the increasing rate of charge injected into box via boundary; $- 2 \Re(O \partial_t \Psi^\dag)$ stands for the generalized work caused by complex scalar field. To understand the term ${\cal T}^{ij}\bar{\nabla}_i\chi_j$, we decompose the energy momentum tensor and $\bar{\nabla}_i\chi_j$ into
\begin{equation}\label{decomptij}
  {\cal T}_{ij}=\mathcal{T}h_{ij}+\pi_{ij},~~\bar{\nabla}_i\chi_j=\theta h_{ij}+\sigma_{ij}+\omega_{ij} \, .
\end{equation}
Here $\pi_{ij}$ stands for the shear stress of cut-off boundary and $\mathcal{T}$ stands for the trace of $\mathcal{T}_{ij}$. The $\theta, \sigma_{ij}$ and $\omega_{ij}$ are the trace, the symmetric traceless and skew-symmetric parts of $\bar{\nabla}_i\chi_j$, respectively, which stand for the the expansion, the shear tensor and rotation tensor of time evolution vector $\chi^i$ respectively. We then see that the first term of the right-side on Eq.~\eqref{ward11} reads
\begin{equation}\label{ward112}
{\cal T}^{ij}\bar{\nabla}_i\chi_j=\mathcal{T}\theta+\pi_{ij}\sigma^{ij} \, .
\end{equation}
Here $\mathcal{T}\theta$ stands for the work caused the variation of area of the boundary and $\pi_{ij}\sigma^{ij}$ stands for the work caused by the shear stress of boundary. The cut-off boundary can be thought of as an elastic membrane and Eq.~\eqref{ward112} then stands for the work cased by the shape-deformation of cut-off boundary. Similar to the Eq.~\eqref{ward21}, we then denote
\begin{eqnarray}\label{ward23}
{\cal P}_{\rm e} \equiv  \bar{\nabla}_j  J^j_{\rm e}
\end{eqnarray}
and interpret it as the energy flux density via the boundary of the box, i.e. the power surface density at the boundary, caused by various works of matters and the deformation of boundary. Specifically, for positive ${\cal P}_{\rm e}$ and ${\cal P}_{\rm Q}$, the box absorbs energy and charge, while for negative ${\cal P}_{\rm e}$ and ${\cal P}_{\rm Q}$, the box loses energy and charge.

\subsection{Complex scalar in frequency domain}
As the first step to study this topic, we now treat the complex scalar as perturbation and neglect the backreaction of scalar field on spacetime geometry. This in fact is ``quasi-static approximation'' and can be applied when discharging current is small enough. In frequency domain, we consider the s-wave mode approximation and adopt the Dirichlet boundary condition at the boundary of the cavity for the scalar field
\begin{equation}\label{source}
\Psi = \te^{-i \omega t} {\cal R}(r)/r \, , \quad {\cal R}|_{r=r_c}= X_0,~~\omega\geq0\, .
\end{equation}
We further leave $\omega$ and $X_0$ as two parameters that we can control.
In frequency domain~[14] and spherically symmetric case, the equation for the complex scalar field will thus reduce into following from
\begin{equation}\label{radial}
\frac{\td^2 {\cal R}}{\td y^2}+ U (y, \omega){\cal R}=0 \, ,
\end{equation}
Here $y$ is the tortoise coordinate and $U (y, \omega)$ is the effective potential which depends on bulk metric and the frequency $\omega$. Near horizon the asymptotical behavior of $ U(y, \omega)$ gives universal behavior $U|_{y \to -\infty} = (\omega-\omega_h)^2$, in which
\begin{equation}
\omega_h \equiv \sigma \mu_h\, , \quad  \mu_h \equiv Q/r_+
\end{equation}
is the electric potential on the horizon. Since the black hole horizon is a one-way membrane, we impose the infalling boundary condition of radial function, namely
\begin{equation}
{\cal R}(y) = X_0 {\cal T}_h \te^{- i (\omega-\omega_h) y},~~y\rightarrow-\infty \, ,
\end{equation}
in which the transmission amplitude ${\cal T}_h$ depends on the parameters related to our black hole battery as well as the frequency $\omega$ and charge $\sigma$. In addition, recall the Eq.~\eqref{respon}, it is easy to observe that ${O}$ is still a periodic function of time, sharing the same frequency with $\Psi$ in frequency domain. We therefore set
\begin{equation}\label{respon2}
O = Z_0 \te^{-i \omega t}/r_c\, .
\end{equation}
Recall the Eq.~\eqref{source}, here we have introduced three coefficients, ${\cal T}_h, X_0$ and $Z_0$, corresponding to the external source and its respond for the complex scalar, but they are not independent. They are further connected by the following conserved Wroskain determinant associated with ${\cal R}$ and its complex conjugate,
\begin{equation}
\left. {\cal R}^\dag \frac{\td{\cal R}}{\td y} - {\cal R}\frac{\td{\cal R}^\dag}{\td y} \right|_{r=r_h} = \left. {\cal R}^\dag \frac{\td {\cal R}}{\td y} -{\cal R}\frac{\td {\cal R}^\dag}{\td y} \right|_{r=r_c} \, .
\end{equation}
giving
\begin{equation}
2|X_0|^2 |{\cal T}_h|^2 (\omega - \omega_h) =i(X_0 Z_0^\dagger-X_0^\dagger Z_0)\, .
\end{equation}
The result can be further simplified if we impose $r_c \gg r_+$. In this case the induced metric of cavity is approximated by the metric of a particle in flat spacetime and gauge potential at the cavity approximately vanishes. Thus, the two ``Ward Identities" of the boundary, the Eq.~\eqref{ward1} and Eq.~\eqref{ward22}, will reduce to
\begin{eqnarray}\label{power1}
 &&{\cal P}_{\rm e} = \frac{2\sigma \omega}{r_c^2} \left( \frac{\omega}{\sigma}-\mu_h \right)|X_0|^2|{\cal T}_h|^2 \, , \cr
 && \\
 &&{\cal P}_{\rm Q} = \frac{2\sigma^2}{r_c^2}\left(\frac{\omega}{\sigma} - \mu_h \right)|X_0|^2|{\cal T}_h|^2 \, . \nonumber
\end{eqnarray}
This recovers energy flux density and charge current density computed from the theory of superradiance~\cite{Brito:2015oca}.

\bibliography{bhbattery1}

\end{document}